\documentstyle[graphicx, psfrag, prd,aps,preprint,tighten]{revtex}
\def\thefootnote{\fnsymbol{footnote}}
\def\bea{\begin{eqnarray}}
\newcommand\eea{\end{eqnarray}}

\begin{document}
\begin{titlepage}
\begin{center}

\hfill LBNL-44439 \\
\hfill UCB-PTH-99/50 \\
\hfill hep-ph/9910478 \\
\hfill \today \\[.3in]

{\large \bf Comment on ``Inflation and flat directions in modular
invariant superstring effective theories''}\footnote{This work was
supported in part by the Director, Office of Energy Research, Office
of High Energy and Nuclear Physics, Division of High Energy Physics of
the U.S. Department of Energy under Contract DE-AC03-76SF00098 and in
part by the National Science Foundation under grant PHY-95-14797.}
\\[.2in]

Mike J. Cai {\em and} Mary K. Gaillard \\[.1in]

{\em Department of Physics, University of California, and \\ Theoretical Physics Group,
50A-5101, Lawrence Berkeley National Laboratory, \\
 Berkeley, California 94720}\\[.5in]

\end{center}

\begin{abstract}
The inflation model of Gaillard, Lyth and Murayama is revisited, with a systematic
scan of the parameter space for dilaton stabilization during inflation.

\end{abstract}
\end{titlepage}

\newpage

\renewcommand{\thepage}{\roman{page}}
\setcounter{page}{2}
\mbox{ }

\vskip 1in

\begin{center}
{\bf Disclaimer}
\end{center}

\vskip .2in

\begin{scriptsize}
\begin{quotation}
This document was prepared as an account of work sponsored by the United
States Government. While this document is believed to contain correct
 information, neither the United States Government nor any agency
thereof, nor The Regents of the University of California, nor any of their
employees, makes any warranty, express or implied, or assumes any legal
liability or responsibility for the accuracy, completeness, or usefulness
of any information, apparatus, product, or process disclosed, or represents
that its use would not infringe privately owned rights.  Reference herein
to any specific commercial products process, or service by its trade name,
trademark, manufacturer, or otherwise, does not necessarily constitute or
imply its endorsement, recommendation, or favoring by the United States
Government or any agency thereof, or The Regents of the University of
California.  The views and opinions of authors expressed herein do not
necessarily state or reflect those of the United States Government or any
agency thereof, or The Regents of the University of California.
\end{quotation}
\end{scriptsize}

\vskip 2in

\begin{center}
\begin{small}
{\it Lawrence Berkeley Laboratory is an equal opportunity employer.}
\end{small}
\end{center}

\newpage
\renewcommand{\thepage}{\arabic{page}}
\setcounter{page}{1}
\def\thefootnote{\arabic{footnote}}
\setcounter{footnote}{0}

\section{Introduction}
The inflation model has proven to be a promising candidate for
describing the early universe.  It offers a very natural and
elegant solution to the horizon and flatness problems in Big
Bang cosmology.  Unfortunately, its success generally relies on fine tuning
some small parameters, and requires one or more
scalar fields (inflatons) to roll slowly down a nearly flat potential.

In principle, a flat potential is not realistic in quantum field
theory.  Any flat potential at tree level will most likely be
destroyed by radiative correction.  However, with the aid of
supersymmetry, such a flat direction may be protected by a
nonrenormalization theorem.  In~\cite{inflat} a model with the
required flatness was constructed, based on the superstring-derived
effective theory of~\cite{moduli}, which utilizes nonperturbative
string effects to stabilize the dilaton in the true
vacuum.  For inflation to be viable, the dilaton must also be
stabilized during inflation. The analytic solution to the
stabilization conditions used in~\cite{inflat} contains an algebraic
error.  In this article, we solve the equations numerically, which
permits a systematic scan of the parameter space for viable solutions.

\section{The model}
The effective potential from orbifold compactification
was presented in \cite{inflat}.  The
K\"ahler potential $K$ and the Green-Schwarz counter term $V_{GS}$ were
taken to be
\bea K &=& G+\ln V + g(V), \quad G=\tilde G + \sum_AX_A, \quad V_{GS}
    =b \tilde G + \sum_A p_A X_A, \nonumber \\
\tilde G &=&\sum_I \tilde G_I, \quad\tilde G_I = -\ln (T_I+\bar T_I -
    \sum_A \left| \Phi_{AI} \right |^2), \quad X_A=\exp \left
    (\sum_I q_I^A \tilde G^I \right) \left| \Phi_A \right|^2,
\eea
where $g(V)$ parameterizes nonperturbative string effects, $V$
is a vector superfield whose scalar component
$V_{\theta=\bar\theta=0}=\ell$ is the dilaton, and $b=30/8\pi^2$ governs
the beta function for $E_8$.  The $T_I$ are the chiral
multiplets containing the moduli.  The $\Phi_{AI}$ are untwisted
sector chiral multiplets, and the $\Phi_A$ are twisted sector
chiral multiplets.  The component Lagrangian was computed in
\cite{moduli}.  Specifically, the scalar potential is given by
\bea 
V&=&\frac{1}{16\ell^2}(\ell g'+1) \left | u(1+b_a \ell) - 4\ell We^{K/2} \right
|^2 - \frac{3}{16} \left | b_a u-4We^{K/2} \right| ^2 \nonumber \\ & &\quad\quad
+ \sum_A \left(\prod_I x_I^{q_I^A} \right)
\frac{|Y_A|^2}{1+p_A \ell} + \sum_I \frac{1}{1+b\ell +
\sum_B(1+p_B\ell)q_i^B X_B} \times \nonumber \\ & & \quad
\Bigg [ \left|A_I (2\xi(t_I) x_I + 1) - e^{K/2} \sum_A \phi_{AI}
W_{AI} \right| ^2
+ x_I \sum_A \left| W_{AI}e^{K/2} + 2\xi(t_I)A_I
\bar\phi_{AI} \right| ^2 \Bigg ]
\eea
where $b_a$ governs the $\beta$-function for the condensing
gauge sector,
\begin{equation}
    A_I=e^{K/2} \left ( \sum_\alpha q_I^\alpha \phi_\alpha W_\alpha -
    W \right ) - \frac{u}{4} (b-b_a),
\end{equation}
and
\begin{equation}
    Y_A=e^{K/2} [W_A +K_A W] +\frac{u}{4}(p_A - b_a)K_A.
\end{equation}

\subsection{Vacuum conditions}
In the true vacuum, all matter fields vanish.  Hence $W=W_a=0$.
Recall that $K_a=\left(\prod_I x_I^{-q_I^a} \right) \bar \phi_a$, which
vanishes in the vacuum as well.  This means
\bea Y_A=0, \quad A_I=-{u\over4}(b-b_a),\quad x_I=t_I+\bar t_I=2
{\rm Re} t_I, \eea
and the scalar potential reduces to
\begin{equation}
    V_0 = \frac{1}{16\ell^2}(\ell g'+1)\left| u(1+b_a\ell)\right|^2 -
    \frac{3}{16}\left |b_a u \right|^2
    +\sum_I \frac{1}{1+b\ell} \left
    |\frac{u}{4}(b-b_a)(2\xi(t_I)x_I+1) \right|^2.
\end{equation}
Minimizing with respect to $t_I$, we obtain $2\xi(t_I)x_I+1=0$.  Therefore,
in the vacuum \footnote{The nonperturbative string effects are
parameterized by two functions $f$ and
$g$,which are related by
$$    \ell g'=f-\ell f', \quad g(\ell=0)=f(\ell=0)=0 $$}
\bea V_0\propto {1\over b_a^2 \ell^2}(\ell g'+1)(1+b_a \ell)^2 -3
    ={1\over b_a^2\ell^2}(f-f'\ell+1)(1+b_a\ell)^2 -3.
\eea 
Now we need to find $f$ such that
\begin{enumerate}
\item  The dilaton is stabilized (${\partial V_0}/{\partial \ell} = 0,
\;{\partial ^2 V_0}/{\partial \ell^2}>0$), and
\item the cosmological constant vanishes ($V_0=0$).
\end{enumerate}
From these two conditions, we arrive at the following constraints:
\bea  &&2(f-f'\ell+1)+\ell^2f''(1+b_a\ell)=0,\quad
    f'''\ell^2(1+b_a\ell)+3b_af''\ell^2 <0,\nonumber \\ &&
    (f-f'\ell+1)(1+b_a\ell)^2-3b_a^2\ell^2=0, \eea
where $\ell=\left< \ell \right>_0$ is the \textit{vev} of
the dilaton in the vacuum.

\subsection{Inflation}
To construct a model of inflation, we make the following
assumptions~\cite{inflat}.
\begin{enumerate}
\item $V^{1/4}\gg \sqrt{u}$.
\item  $W\sim 0$.
\item $W_\alpha=0$, except for $\alpha =C3$, which is in the untwisted
sector.
\item All matter field $vev$'s are negligible. \end{enumerate}
 Then the scalar potential during inflation is
\begin{equation}
    V_i=\frac{\ell e^g}{(1+b\ell)x_1x_2}|W_{C3}|^2.
\end{equation}
It is expected that $W_{C3}$ has a power law dependence on the
dilaton, which will be discussed later.  The dilaton dependence of
$V_i$ can be written as
\begin{equation}
    V_i = \frac{\ell^d e^g}{(1+b\ell)}.
\end{equation}
Once again, we need to stabilize the dilaton.  This time, there is
an extra constraint.  That is, the dilaton \textit{vev} during
inflation is located in the domain of attraction of the true
vacuum.  Dilaton stabilization equations are
\bea f-f'\ell+d-{b\ell\over 1+b\ell} =0,\quad
f''+ {1\over b\ell(1+b\ell)^2}<0.\eea

\subsection{Summary of the equations for dilaton stabilization}
The stabilization equations are most simply expressed in terms
of the rescaled dilaton field $ \zeta = b \ell. $
In terms of this variable they take the following form.
\subsubsection{Vacuum: $\zeta = b \left< \ell\right>_0$}
\bea f''+\frac{6\gamma^2}{(1+\gamma \zeta)^3} =0 \label{1st eqn},\quad
    f-f'\zeta+1-\frac{3\gamma^2\zeta^2}{(1+\gamma \zeta)^2} =0,\quad
    f'''-\frac{18\gamma^3}{(1+\gamma \zeta)^4}<0,\eea
where \bea b=\frac{30}{8\pi^2}, \quad \gamma=b_a/b. \eea

\subsubsection{Inflation: $\zeta=b\left<\ell\right>_i$}
\bea f-f'\zeta+d-\frac{\zeta}{\zeta+1} =0,\quad
     f'' +\frac{1}{\zeta(1+\zeta)^2} <0. \label{last eqn}
\eea
For simplicity, we will use only the two leading terms for the
nonperturbative parameters \cite{nonpert}.
\begin{equation}
    f(\zeta)=B\left (1+ A\sqrt{\frac{a}{\zeta}} \right )
    e^{-\sqrt{a/\zeta}},
\end{equation}
where $A$, $B$ and $a$ are adjustable parameters.
As opposed to the previous equations, all derivatives that appear in these
equations are with respect to the rescaled dilaton $\zeta$.

\section{Phenomenological constraints on the parameters}
\subsection{The parameter $\mathbf{\gamma}$}
The effective gauge coupling at the string scale is
$g^{-2}=(f+1)/2\ell$.  Recall that the gravitino mass is given by
\begin{equation}
    M_{\tilde{G}} = \frac{1}{4} b_a \left| \left< \bar{\lambda}
    \lambda \right> \right |, \quad M_p=1,\end{equation}
where $M_p$ is the reduced Planck mass: $M_p = (8\pi G_N)^{-2}$.
To establish the observed hierarchy, we want $M_{\tilde{G}}\sim 1TeV$.
 This determines the supersymmetry (SUSY) breaking scale:
\bea M_{\tilde{G}}=\frac{1}{4}b_a \Lambda^3/M_p^2 \sim 10^3 GeV, \quad
     \Lambda \sim 10^{14} GeV, \eea
assuming $b_a \sim O(0.1)$.  If SUSY is broken by a condensate, the
renormalization group equation (RGE)
tells us the scale $\Lambda$ at which the gauge interaction becomes strong;
in the leading log approximation
\begin{equation}
    \mu \frac{\partial g}{\partial \mu} = -{3\over2}b_a g^3,\quad
\Lambda =M_p \exp \left(-1/3b_a g^2 \right).
\end{equation}
For $\mu = \Lambda \sim 10^{14} GeV$, $3b_ag^2 \sim 0.1$.  This
relates $\gamma$ to $g^2$:
\begin{equation}
    \gamma=\frac{b_a}{b} \sim .03 \frac{f+1}{2 \zeta}.
\end{equation}

\subsection{The parameter ${\bf d}$}
The D-term in the scalar potential contains a
Fayet-Illiopoulos term:
\begin{equation}
    V_D=\frac{g^2}{2}\left (\sum q_n K_n \phi_n + \xi_D\right) ^2
\end{equation}
where $K_n\propto \bar\phi_n$, and $\xi_D \propto \ell$.  This leads
to a \textit{vev} $\left< \phi_n \right> \propto \ell^{1/2}$.  This
will in turn induce other \textit{vev}'s of the form $\left
<\phi_n \right > \propto \ell^{-1}$  \cite{inflat}.
The superpotential in general has a power series expansion in all
the matter fields.  Since $V_i \propto \left|W_{C3}\right|^2$, we
conclude that $d$ is an {\it integer}, which may take on
negative values.

\section{Results}
The equations (\ref{1st eqn})-(\ref{last eqn}) are solved self-consistently based
on two input parameters: $d$ and the gauge coupling $g$.  The
upper bound of $d$ is determined by the inflation equation.  In
this case, there is no solution for $d \geq 2$.  The lower bound
of $d$ is determined by the requirement that the dilaton remains
in its domain of attraction.  In the following table, the variables
are defined as follows:
\begin{enumerate}
\item $g_{max}^2$:  the maximum value of $g^2$ such that the
equations have solutions.
\item $\ell_0$:  \textit{vev} of the dilaton in vacuum.
\item $\ell_i$:  \textit{vev} of the dilaton during inflation.
\end{enumerate}

The RGE extrapolation of low energy couplings in the context of
of the Minimal Supersymmetric Standard Model (MSSM) gives $g^2\sim .5$
at a scale of about $10^{16} GeV$.  Unification at the string scale,
$\mu_s = g$ in reduced Planck mass units, can be achieved~\cite{dienes}
by adding additional matter fields. This increases $g^2$, in some cases
to a value as high as $g^2 \approx 1$.
Hence we conclude that $d=1$, $d=0$ and $d=-1$ are
candidates for a realistic model. 

A typical solution is plotted here.  Notice that in the scalar
potential, an overall normalization proportional to the gaugino
condensate is not included.

\begin{figure}
\psfrag{V0}{$V_0$}
\psfrag{vacuum}{scalar potential in the vacuum}
\psfrag{bl}{$b \left< \ell \right>_0$}
\includegraphics[width=.9 \textwidth]{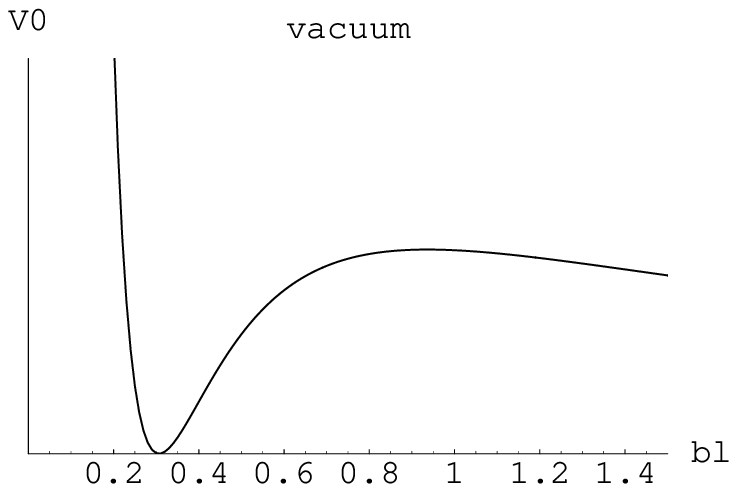}
\caption{Input parameters:  $d=1$, $g^2_{string}=1.46$}
\end{figure}

\begin{figure}
\psfrag{Vi}{$V_i$}
\psfrag{inflation}{scalar potential during inflation}
\psfrag{bl}{$b \left< \ell \right>_i$}
\includegraphics[width=.9 \textwidth]{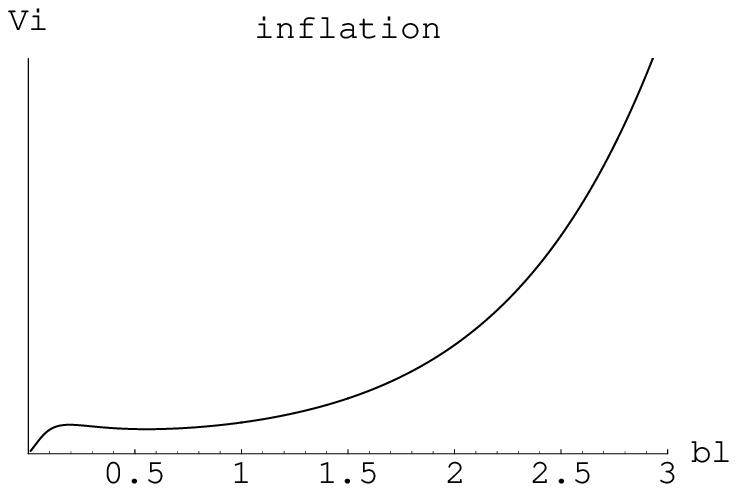}
\caption{Input parameters:  $d=1$, $g^2_{string}=1.46$}
\end{figure}

\begin{table}
\begin{tabular}{ccccccc}
$d$ &   $g_{max}^2$ &   $\ell_0$&   $\ell_i$&   $A$    & $B$    & $a$\\
$1$ &   $2.3$      &   $1.16$  &   $2.76$  &   $-.66$ & $9.07$ & $1$\\
$0$ &   $.73$       &   $.69$   &   $2.10$  &   $-.36$ & $20.6$ & $1$\\
$-1$&   $.68$       &   $1.01$   &   $3.51$  &   $-.063$ & $39.2$ & $3$\\
$-2$&   $.15$       &   $.22$   &   $.89$  &   $-.069$ & $39.0$ & $.65$\\
\end{tabular}
\caption{Parameters for different values of $d$}
\end{table}
\end{document}